\documentclass[preprint]{aastex}
\shortauthors{CHUNG $\&$ Park}
\shorttitle{CENTRAL PERTURBATIONS BY PLANETS IN BINARIES}

\newcommand{\thetae}{\theta_{\rm E}}
\newcommand{\thetaeone}{\theta_{\rm {E,1}}}
\newcommand{\sbb}{s_{\rm b}}
\newcommand{\sbhat}{{\hat s}_{\rm b}}
\newcommand{\qb}{q_{\rm b}}
\newcommand{\spp}{s_{\rm p}}
\newcommand{\spphat}{{\hat s}_{\rm p}}
\newcommand{\qp}{q_{\rm p}}
\newcommand{\delxib}{\Delta \xi_{\rm b}}
\newcommand{\delxip}{\Delta \xi_{\rm p}}

\begin{document}

\title{Properties of Microlensing Central Perturbations by Planets in Binary Stellar Systems under the Strong Finite-Source Effect}

\author{
Sun-Ju Chung and Byeong-Gon Park
}
\affil{
Korea Astronomy and Space Science Institute, Hwaam-Dong,
Yuseong-Gu, Daejeon 305-348, Korea; sjchung@kasi.re.kr, bgpark@kasi.re.kr
}


\begin{abstract}
We investigate high-magnification events caused by planets in wide binary stellar systems under the strong finite-source effect, where the planet orbits one of the companions.
From this investigation, we find that the pattern of central perturbations in triple lens systems commonly appears as a combination of individual characteristic patterns of planetary and binary lens systems in a certain range where the sizes of the caustics induced by a planet and a binary companion are comparable, and the range changes with the mass ratio of the planet to the planet-hosting star.
The inside and outside edge regions of a circle with a radius corresponding to that of the source star and its center located at the center of the caustic, show the binary-lensing pattern, while the outside region of the circle shows the planetary-lensing pattern.
Specially, we find that because of this central perturbation pattern, the characteristic feature of high-magnification events caused by the triple lens systems appears in the residual from the single-lensing light curve despite the strong finite-source effect, and it is discriminated from those of the planetary- and binary-lensing events and thus can be used for the identification of the existence of both planet and binary companion.
This characteristic feature is a simultaneous appearance of two features.
First, double negative-spike and single positive-spike features caused by the binary companion appear together in the residual, where the double negative spike occurs at both moments when the source enters and exits the caustic center and the single positive spike occurs at the moment just before the source enters into or just after the source exits from the caustic center.
Second, the magnification excess before or after the single positive-spike feature is positive due to the planet, and the positive excess has a remarkable increasing or decreasing pattern depending on the source trajectory.
\end{abstract}

\section{INTRODUCTION}

The microlensing signal of a planet is a short-duration perturbation on the smooth standard light curve of the primary-induced lensing event that occurred on a background source star.
To detect extrasolar planets using microlensing, survey and follow-up observations are now being carried out toward the Galactic bulge field.
The survey observations (OGLE: Udalski 2003, MOA: Bond et al. 2002) monitor a large area of sky and alert ongoing events by analyzing data in real time, while the follow-up observations ($\mu$FUN: Dong et al. 2006, PLANET: Albrow et al. 2001, RoboNet: Burgdorf et al. 2007) intensively monitor the alerted events.

For a planetary-lensing event, the perturbation of the lensing light curve is induced by two sets of disconnected caustics that are composed of the central and planetary caustics.
The perturbation induced by the central caustic always occurs close to the peak of the lensing light curve, while the perturbation induced by the planetary caustic can occur at any part of the light curve.
The lensing event caused by the central caustic becomes a high-magnification event with central perturbation and this event is very sensitive for the detection of a planet \citep{griest98}.
The current follow-up observations thus focus on high-magnification events.
However, the lensing event by a very close or a very wide binary can also produce central perturbation of the high-magnification event.
Fortunately, \citet{han08a} found that high-magnification events with a double-peak structure caused by a planet and a binary companion can be immediately distinguished by the shape of the interpeak region of the light curves.

The size of the source star becomes important in high-magnification events because the source star passes very close to the central caustic.
In planetary-lensing events where the source diameter is considerably larger than the central caustic and thus the finite-source effect is strong, the central perturbation is greatly buried and the resulting light curve appears like that of a single-lensing event induced by the primary star.
Thus, planetary- and binary-lensing events with a double-peak structure affected by the strong finite-source effect cannot be distinguished any more by the shape of the interpeak region of the light curves.

High-magnification events with strong finite-source effects have been recently reported by \citet{subo09} and \citet{janczak10}.
They reported that, on casual inspection, these events appear as single-lensing events with pronounced finite-source effects, but a more detailed analysis reveals they are planetary-lensing events with a buried signature of the planet.
\citet{hankim09} investigated the planetary-lensing signals of high-magnification events under strong finite-source effects.
They found that the characteristic features of planetary-lensing events commonly appear in the residuals and thus they can be used for the diagnosis of the existence of the planet.
However, they noted that the characteristic residual features can be also produced by very close or very wide binary-lensing events and thus the existence of these features does not necessarily confirm the existence of the planet.
We expect that many more high-magnification events with buried signatures of the planet and binary companion will be detected by future observations with high cadence monitoring including the ground-based observations (e.g., Korea Microlensing Telescope Network (KMTNet) Project: B.-G. Park 2009, private communication) and space-based observations (e.g, {\it Microlensing Planet Finder} ({\it MPF}): Bennett et al. 2004).
Therefore, it is very important to distinguish planetary- and binary-lensing events with the buried signature.
\citet{han09} found that some planetary- and binary-lensing events with the buried signature can be distinguished by the difference of the characteristic features of the individual residual patterns.

For triple-lensing events caused by planets in binary stellar systems, where the sizes of the caustics induced by a planet and a binary companion are similar to each other, the signatures of both planet and binary companion can show up in the lensing light curves, and thus it is possible to detect planets in binary stellar systems \citep{lee08}.
However, if the events are strongly affected by finite-source effects, both signatures are greatly washed out and thus do not obviously appear in the light curves.
After all, triple-lensing events are very difficult to be identified as events caused by planets in binary systems.
Fortunately, since current follow-up observations have a high photometric precision ($\sim 1 \%$) at the peaks of high-magnification events and it is expected that future observations would have more high photometric precision, it could be feasible for the identification of the planetary signatures in binary stellar systems affected by the strong finite-source effect.
In this paper, we investigate high-magnification events caused by planets in binary stellar systems to find out whether the signatures of both planet and binary companion can be identified despite the strong finite-source effect.

The paper is organized as follows. In Section 2, we briefly describe the central caustics caused by a planet and a binary companion. In Section 3, we investigate the central perturbation patterns of planets in binary systems under the strong finite-source effect.
We also compare the residual patterns of the events caused by the triple lens systems with those of planetary- and binary-lensing events. We summarize the results and conclude in Section 4.

\section{CENTRAL CAUSTIC}

To be identified as a planet in a binary stellar system, the binary companion should be located within an optimal separation range, in which the sizes of the caustics caused by the planet and binary companion are comparable \citep{lee08}.
This is because the two comparable caustics induce a similar amount of perturbation in a region around the primary star and thus the signatures of both planet and companion appear well in the lensing light curve.
However, if the two caustics are smaller than the diameter of the source star, the two signatures in the lensing light curve will be very washed out by the strong finite-source effect, and thus the light curve of the event will seemingly appear to be that of the single-mass lensing event of the primary star alone.

For a very widely separated binary, the size of the caustic caused by a binary companion is represented by
\begin{equation}
\delxib \simeq {4\gamma \over \sqrt{1 - \gamma}} = {4\qb \over \sbhat^2}\left (1 + {\qb \over 2\sbhat^2} \right ) ; \quad
\gamma = {\qb \over \sbhat^2},\ \ \ \qb = {m_{2} \over m_{1}},
\end{equation}
where $\gamma$ is the shear induced by the binary companion, $\sbb$ is the projected separation between the binary components normalized by the Einstein radius corresponding to the total mass of the lens system, $\thetae$, $m_1$ and $m_2$ are the masses of the primary and companion stars, respectively \citep{lee08}.
Here, the notation with the $``hat"$ represents the length scale in units of the Einstein ring radius of the primary, $\thetaeone = \thetae [m_{1}/(m_{1}+m_{2})]^{1/2}$.

On the other hand, the size of the central caustic caused by a planet is given by
\begin{equation}
\delxip \simeq {4\qp \over {(\spp - 1/\spp)^2}}\ ; \quad \qp = {m_{p} \over m_{1}},
\end{equation}
where $\spp$ is the projected primary-planet separation normalized by $\thetae$ and $m_p$ is the planet mass \citep{chung05}.
For the planetary lensing with $\qp \ll 1.0$, the Einstein radius of the lens system is approximated to the Einstein radius of the primary, i.e., $\thetae \sim \thetaeone$, and thus $\spp \sim \spphat$.

\section{CENTRAL PERTURBATION PATTERN}

\subsection{Excess Map}

We consider the geometry of triple lens systems composed of a planet and binary stars, where the planet is orbiting around the primary star, and the binary companion is widely separated from the primary star.
To investigate the perturbation pattern of high-magnification events caused by the triple lens systems under the strong finite-source effect, we construct magnification excess maps of the triple lens systems.
The magnification excess is defined by
\begin{equation}
\epsilon = {A - A_{0} \over {A_0}}\ ,
\end{equation}
where $A$ and $A_0$ are the triple- and single-lensing magnifications, respectively.
For limb darkening, we adopt a brightness profile for the source star of the form
\begin{equation}
{I(\theta)\over{I_0}} = 1 - \Gamma \left(1-{3\over{2}}{\rm cos}\theta \right ) - \Lambda \left(1-{5\over{4}}{\rm cos}^{1/2}\theta \right ) ,
\end{equation}
where $\Gamma$ and $\Lambda$ are the linear and square root coefficients and $\theta$ is the angle between the normal to the surface of the source star and the line of sight.
Following \citet{han09}, we fix $\Gamma = -0.46$ and $\Lambda = 1.1$.

Figure 1 shows the magnification excess maps of planets in binary stellar systems for various planet/primary mass ratios and size ratios of the caustics induced by a planet and a binary companion.
The dashed circle has a radius corresponding to that of the source and is located at the center of the caustic.
In each map, the regions with blue and red-tone colors represent the areas where the excess is negative and positive, respectively.
The color in Figure 1 changes into darker scales when the excess is $|\epsilon| = 2\%,\ 4\%,\ 8\%,\ 16\%,\ 32\%$, and $64\%$, respectively.
In the map, the lensing parameters $(\qb,\spp)$ are $(0.5,1.23)$ and the position angle of the planet measured from the binary axis is $50^\circ$.
We assume that the source is a subgiant with a radius of $1.76\ R_\odot$ and located at $D_{\rm S} = 8$ kpc, as was the case for the event MOA-2007-BLG-400 \citep{subo09}, and the mass and distance of the lens are $m = 0.3\ M_\odot$ and $D_{\rm L} = 6$ kpc, respectively.
Then, the corresponding angular Einstein radius is $\thetae = 0.32\ \rm{mas}$, and the source radius normalized to the Einstein radius is $\rho_{\star} = \theta_\star/\thetae = (R_\star/D_{\rm S})/\thetae = 0.0032$.

From the map, we find that the pattern of the central perturbations of the triple lens systems affected by the strong finite-source effect appears as a combination of the characteristic patterns of planetary and binary lens systems in a certain range where the sizes of the caustics induced by a planet and a binary companion are comparable, and the range varies depending on the planet/primary mass ratio, i.e.,
\begin{equation}
\sqrt{\qp \over {10^{-3} }} \lesssim {\delxip \over \delxib} \lesssim 1.8\sqrt{\qp \over{10^{-3}}}\\ \\.
\end{equation}
In this range, the inside and outside edge regions of the dashed circle are dominated by the binary companion and the outside region of the circle is dominated by the planet.
Both the inside and outside regions of the circle in the ranges where ${\delxip / \delxib} < \sqrt{\qp /{10^{-3}}}$ and ${\delxip / \delxib} > 1.8 \sqrt{\qp / {10^{-3}}}$ are dominated by the companion and planet, respectively.
As a result, the inside and outside edge regions of the dashed circle show the binary-lensing pattern that individually forms a complete negative-excess annulus and four localized arc-shaped positive excesses inside and just outside the circle \citep{han09}, whereas the outside region of the circle shows the planetary-lensing pattern that forms the positive and negative excesses around the cusps and fold caustics as the shape of the central caustic induced by the planet \citep{hankim09}.

Figure 2 shows the light curves and residuals from the single-lensing event resulting from the source trajectories presented in Figure 1.
In the upper part of each panel, solid and dashed curves are the light curves of the triple- and single-lensing events, respectively.
The lower part shows the residual from the single-lensing light curve.
As shown in Figure 2, all the perturbations induced by a planet and a binary companion in the lensing light curves are greatly washed out by the strong finite-source effect, but many triple-lensing events in the figure still have the characteristic features of both planetary- and binary-lensing events in the residuals.
First, double negative-spike and single positive-spike features caused by the negative-excess annulus and localized arc-shaped positive excess simultaneously appear in the residual, where the negative spike occurs at both moments when the source enters and exits the caustic center, and the positive spike occurs at the moment just before the source enters into or just after the source exits from the caustic center.
The features are the binary-lensing features.

Second, the next excess of the positive spike that occurs just after the exit of the source from the caustic center is positive by the cusp of the central caustic with positive excess induced by the planet, and this is the planetary-lensing feature in the residual of the triple-lensing event.
For binary-lensing events, the excesses after the positive spike in the residual are mostly negative, because the localized arc-shaped positive excesses causing the positive spike are formed in the direction of the fold caustic with negative excess.
The positive excess after the positive spike also shows a noticeable increasing pattern.
However, if the source trajectory is opposite, then the positive spike will appear just before the entrance of the source into the caustic center, and the excess before the positive spike will show a noticeable decreasing positive-excess pattern.
We note that the positive spike or the negative spike means a well-resolved dip in the observed residual as the definition of \citet{han09} for the spike.

\subsection{Comparion with Planetary- and Binary-Lensing Features}

To find out whether above triple-lensing features in the residual can be distinguished from the characteristic features of planetary- and binary-lensing events, we compare the residual patterns of the triple-lensing events with those of the two kinds of events.

Figure 3 shows the magnification excess maps of planet-binary, stellar binary, and planetary lens systems together with the residuals resulting from the source trajectories presented in the individual maps.
Here the planet-binary and stellar binary lens systems are composed of wide binary stars with and without a planet, respectively, and the planetary lens system is composed of a star and a planet with $2\rho_{\star} \simeq \delxip $.
Trajectories I and II in the planet-binary map of Figure 3 are representative paths where the residual patterns are possible to be distinguished from those of planetary and binary events because of the formation of the positive excess region after a local arc-shaped positive excess.
In this planet-binary lensing map, other residual patterns except two kinds of residual patterns like those of trajectories I and II can be generally produced by binary-lensing events and thus are difficult to be discriminated from those of binary-lensing events.
This is because central perturbation patterns by planet-binary lens systems are similar to those by binary lens systems without planets.
We thus consider only the two residual patterns of planet-binary (triple) lensing events for the comparison with those of planetary- and binary-lensing events.

In the panels for trajectories I and II of Figure 3, both double negative-spike and single/double positive-spike features induced by the binary companion appear prominently in the residuals of the triple- and binary-lensing events.
However, the excess patterns of the triple-lensing events before or after the positive spike are different from those of the binary-lensing events.
For the triple-lensing event of trajectory I, the excess before the positive spike is deeper than the double negative-spike feature because the fold caustics with negative excess induced by the planet and binary companion are superposed each other and thus the negative excess around the superposed fold caustic becomes stronger.
The depth of the double negative-spike feature for binary-lensing events is always deeper than or similar to that of the negative excess before or after the positive-spike feature.
This is because the negative-excess annulus in the dashed circle is featured more strongly or similar to the negative excess outside the circle, as shown in the binary-lensing map of Figure 3.
Therefore, the binary-lensing events cannot produce the triple-lensing feature of the trajectory I.
In addition, the excesses after the positive spike for the triple-lensing events are all positive unlike those of the binary-lensing events.
This is because the central caustic induced by the planet has been formed between the fold caustics with negative excess induced by the binary companion.
By virtue of these triple-lensing features appearing before and after the positive spike, the triple-lensing events can be discriminated from the binary-lensing events.

Some binary-lensing events, however, can mimic the triple-lensing feature that produces the positive excess after the positive spike.
In the case of the residuals of the binary-lensing events for trajectories III and IV in Figure 3, the excesses after the positive spike are all positive such as the triple-lensing case, but the pattern of the excesses is different from that of the triple-lensing events.
For the binary-lensing events, the positive excess after the positive spike shows a rather flat pattern, because the source should pass the region with a weak and almost unchanging level in order to produce the triple-lensing pattern (see the trajectories III and IV in the binary-lensing map of Figure 3).
Therefore, this flat positive-excess pattern can be resolved from the triple-lensing pattern with a remarkable increasing or decreasing positive-excess.

For planetary-lensing events, the double negative-spike feature or the single/double positive-spike feature can occur in the residuals, as shown in Figure 3 of \citet{han09}, but it is very hard to simultaneously produce the two features due to the property of the planetary-lensing perturbation pattern (see Figure 1 of \citet{han09}).
Moreover, the excesses after the positive spike in the residuals of the planetary-lensing events are mostly negative due to the same reason as the binary-lensing case mentioned in Section 3.1.
However, the limited planetary-lensing events where $2\rho_{\star} \simeq \delxip $ (e.g., the right panel of Figure 3) show similar behaviors as the residual patterns of triple-lensing events.
As shown in the residual of the planetary-lensing event for trajectory V in Figure 3, the planetary-lensing event produces the double negative- and positive-spike features as well as a deep negative-excess pattern before the positive spike.
The excess after the positive spike also has an increasing positive-excess pattern, and thus this residual pattern is equivalent to that of the triple-lensing event for the trajectory I.

On the other hand, the planetary-lensing event for trajectory VI produces both the double negative-spike and single positive-spike features, but the positive excess after the positive spike has a flat pattern like the binary-lensing events.
As a result, we find that the simultaneous appearance of the following features of triple-lensing events caused by planets in binary stellar systems under the strong finite-source effect is discriminated from the characteristic features of planetary- and binary-lensing events, and thus it can be used for the diagnosis of the identification of the planets in the binary systems.
\begin{enumerate}
\item
Double negative-spike and single positive-spike features appear together in the residual of the lensing event.
\item
The excess before or after the single positive spike in the residual is positive, and the positive excess has a remarkably increasing or decreasing pattern depending on the source trajectory.
\end{enumerate}
Here, the first and second features indicate the existence of the binary companion and planet, respectively.
Since the two features of the triple-lensing events directly depend upon the central perturbation of the triple-lensing systems, they appear within the range of Equation (5).
Figure 2 shows that the events caused by the triple lens systems within the range of Equation (5) have above two features in the residuals.
Considering that the photometric error of current follow-up observations reaches $\sim 1\%$ at the peaks of high-magnification events, the spikes with deviation $\gtrsim 3\%$ can be readily detected \citep{han09}.

However, we note that in triple-lensing systems where the planets are located on or around the axes that are parallel with and normal to the binary axis, it is difficult to simultaneously produce above two features even though the lensing systems are in the range of Equation (5).
This is because the positive excess region that is formed after a localized arc-shaped positive excess and thus produces the second feature of the two features, is not formed in these lens systems.
We also note that in the limited case of a host star with two planets, where the sizes of the caustics induced by the individual planets are similar and the diameter of the source is bigger than the caustics, the lensing event can also produce simultaneously above two features due to the similarity of the central perturbation pattern.
Therefore, a more detailed analysis is required to surely identify whether the two features are caused by a planet and binary stars or a star with two planets.

\section{CONCLUSION}
We have investigated high-magnification events caused by planets in binary stellar systems under the strong finite-source effect, especially for wide-separation binaries with low mass planets.
From this study, we found that the central perturbation patterns of the triple lens systems commonly appear as the combination of the characteristic perturbation patterns of planetary and binary lens systems in a certain range that changes as the planet/primary mass ratio.
The inside and outside edge regions of a circle with a radius corresponding to that of the source and its center located at the caustic center show the binary-lensing pattern, while the outside region of the circle shows the planetary-lensing pattern.
In particular, we found that due to this central perturbation pattern, the characteristic feature of the high-magnification events caused by the triple lens systems appears in the residual from single-lensing light curve, and it is distinguished from those of the planetary- and binary-lensing events and thus can be used for the diagnosis of the identification of the existence of both planet and binary companion.
This characteristic feature is a simultaneous appearance of the following two features.
First, double negative-spike and single positive-spike features caused by the binary companion appear together in the residual, where the double negative spike occurs at both moments when the source enters and exits the caustic center, and the single positive spike occurs at the moment just before the source enters into or just after the source exits from the caustic center.
Second, the excess before or after the single positive spike in the residual is positive due to the planet, and the positive excess has a remarkably increasing or decreasing pattern depending on the source trajectory.

We thank I. A. Bond for making helpful comments.

\begin{figure}[t]
\epsscale{1.0}
\plotone{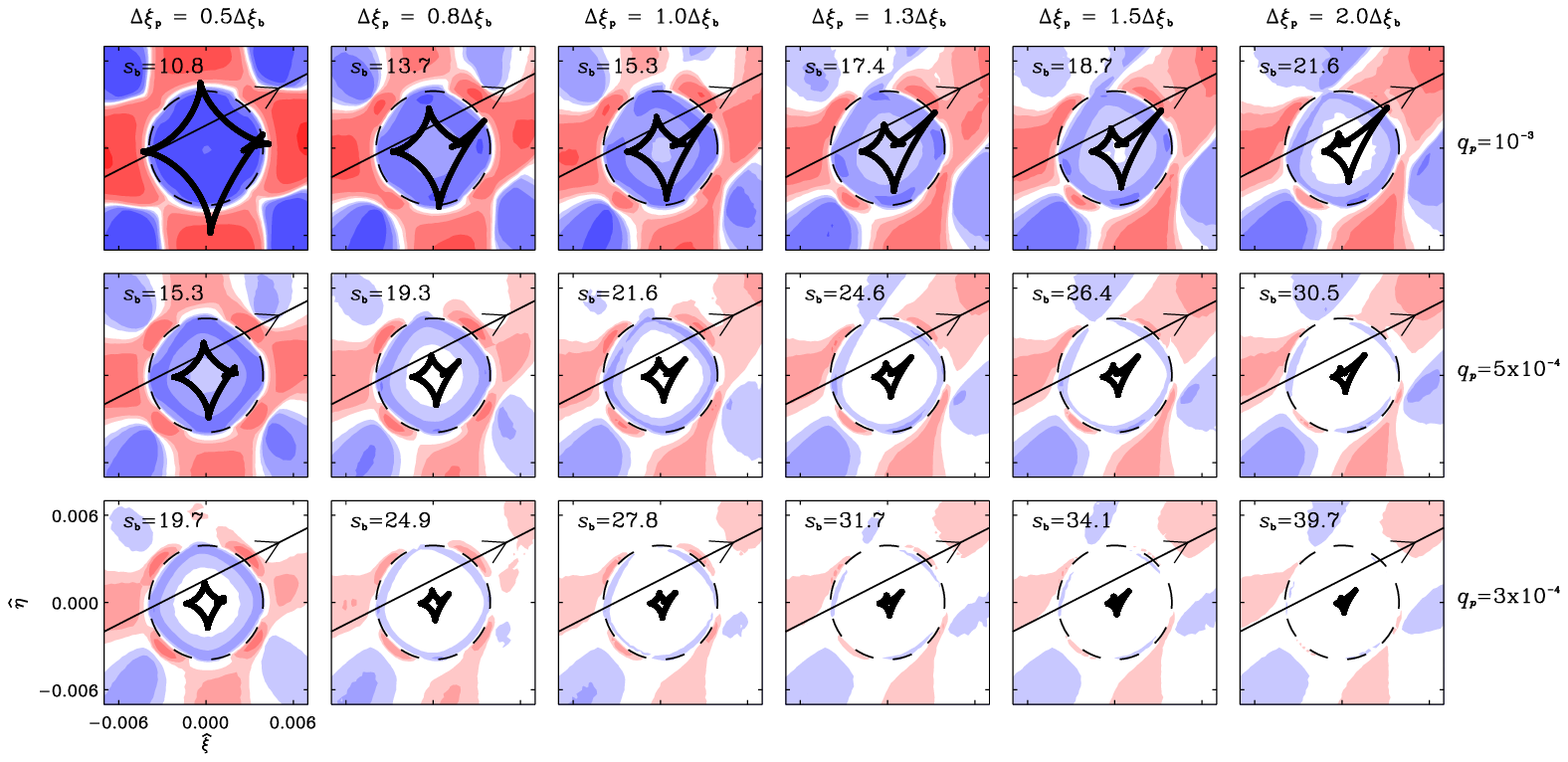}
\caption{\label{fig:one}
Magnification excess maps of planets in binary stellar systems with various planet/primary mass ratios and size ratios of the caustics induced by a planet and a binary companion.
The coordinates ($\hat{\xi},\hat{\eta}$) represent the axes that are parallel with and normal to the binary axis and are centered at the effective position of the primary star.
Here the notation with the hat represents the length scale normalized by the Einstein radius of the primary, $\thetaeone$.
In this map, the companion/primary mass ratio and projected primary-planet separation are $(\qb, \spp) = (0.5, 1.23)$, the position angle of the planet measured from the binary axis is $50^\circ$, $\sbb$ is the projected separation between the binary components in units of the Einstein radius corresponding to the total mass of the lens system, $\thetae$, $\qp$ is the planet/primary mass ratio, and $\delxip$ and $\delxib$ are the sizes of the caustics induced by the planet and binary companion, respectively.
The color changes into darker scales when the excess is $|\epsilon| = 2\%,\ 4\%,\ 8\%,\ 16\%,\ 32\%$, and $64\%$, respectively.
The dashed circle represents the moment when the source star is located at the center of the caustic.
The straight lines with arrows represent the source trajectories.
}\end{figure}

\begin{figure}[t]
\epsscale{1.0}
\plotone{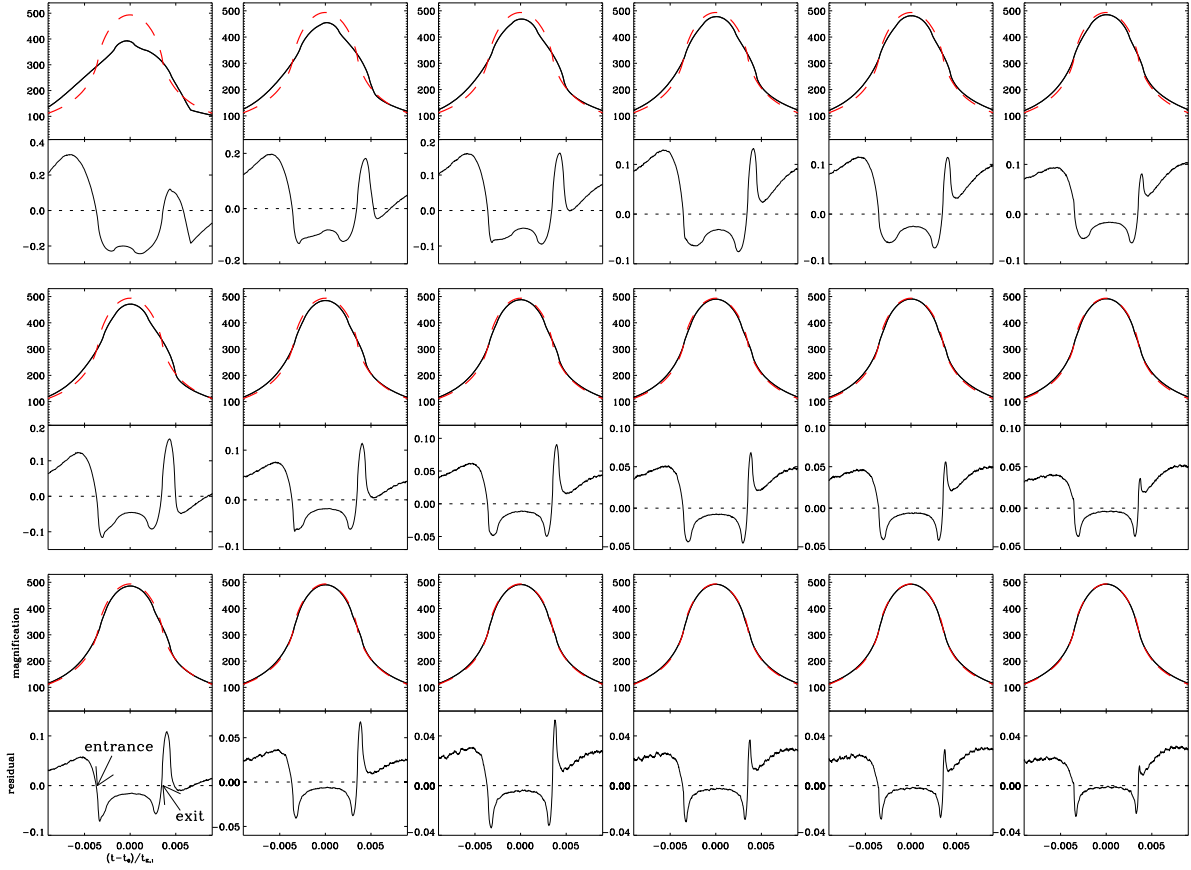}
\caption{\label{fig:two}
Light curves for the source trajectories presented in Figure 1.
In the upper panel, solid and dashed curves represent the light curves of the triple- and single-lensing events, respectively.
The lower panel shows the residuals from the single-lensing light curve.
In the lower panel, the horizontal line indicates the magnification excess of $|\epsilon|=0.0$.
The arrows in the left bottom panel represent the moments when the source enters and exits the dashed circle located at the caustic center in Figure 1.
}\end{figure}

\begin{figure}[t]
\epsscale{1.0}
\plotone{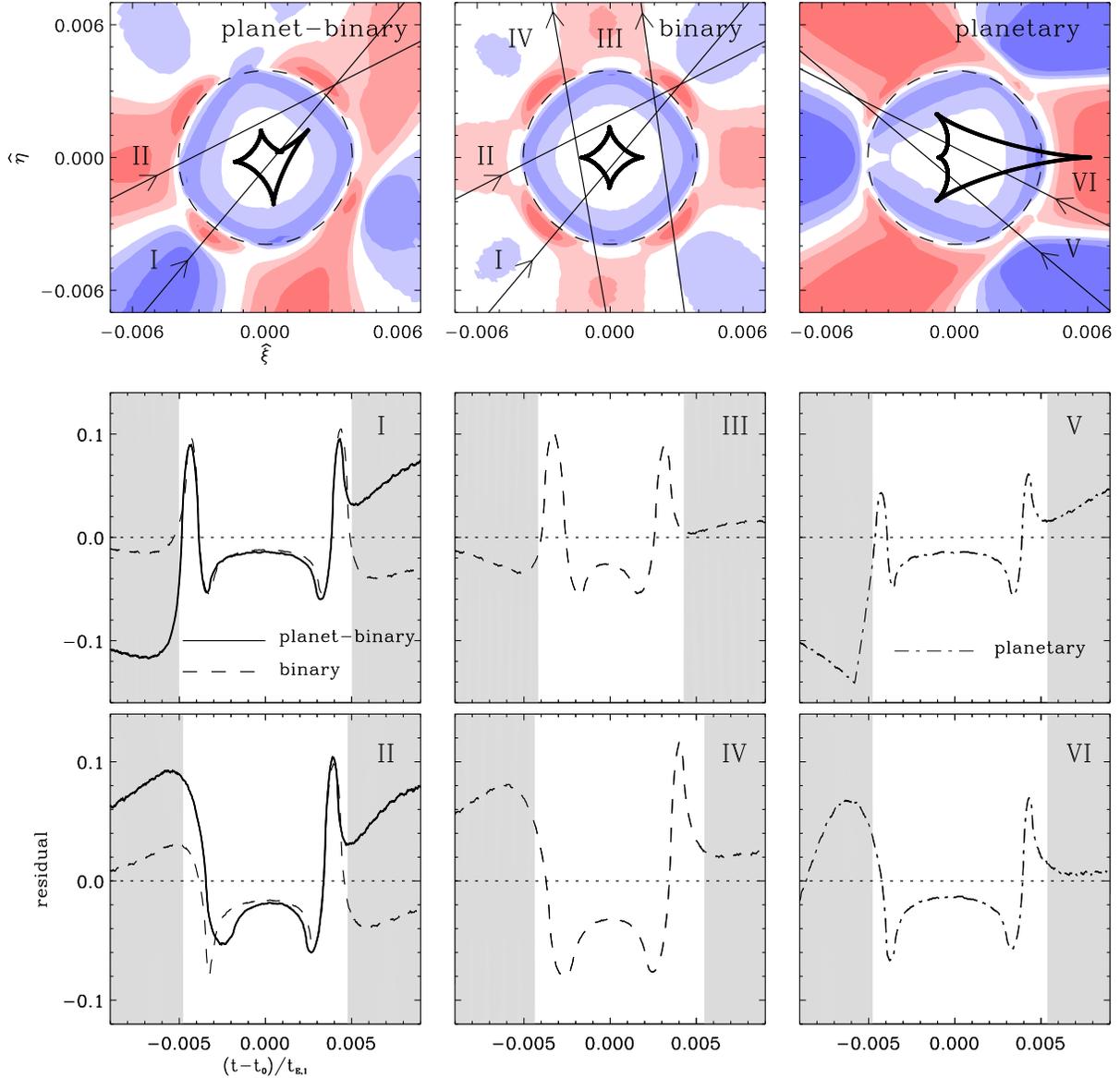}
\caption{\label{fig:three}
Magnification excess maps of planet-binary, stellar binary, and planetary lens systems together with the residuals resulting from the source trajectories presented in the individual maps.
Solid, dashed, and dashed-dot curves represent the residuals of the planet-binary (triple), binary-, and planetary-lensing events, respectively.
Shaded regions represent the perturbations occurring over the outside edge of the dashed circle.
}\end{figure}


\end{document}